# Darkverse - a new Dark Web?

Raymond Chan, Benjamin W.J. Kwok, Adriel Yeo, Kan Chen, Jeannie S. Lee
Singapore Institute of Technology, Singapore
{Raymond.Chan,Benjamin.Kwok,Adriel.Yeo,Kan.Chen,Jeannie.Lee}@singaporetech.edu.sg

## ABSTRACT

The "Darkverse" could be the negative harmful area of the Metaverse; a new virtual immersive environment for the facilitation of illicit activity such as misinformation, fraud, harassment, and illegal marketplaces. This paper explores the potential for inappropriate activities within the Metaverse, and the similarities between the Darkverse and the Dark Web. Challenges and future directions for investigation are also discussed, including user identification, creation of privacy-preserving frameworks and other data monitoring methods.

## 1 INTRODUCTION

The Metaverse is envisioned as a Virtual Reality (VR) space that allows users to interact with one another in a computer generated environment. Several multi-user collaborative applications and platforms have launched or gained popularity in recent years [1–3, 11, 16, 17], such as Meta Horizon Worlds [14], Fortnite [8], Minecraft [15], VR Chat [19] and Decentraland [6]. On these platforms, users communicate, play, socialise and work through digital avatars within an immersive virtual space. However, these platforms are also liable to misuse and abuse, due to the embodied virtual space, mirroring environments and situations in reality. Having a long past history in social media platforms (e.g. Facebook, Twitter and Instagram), scams, phishing, trolling and harassment are now also problems on VR gaming platforms.

The Dark Web refers to services and applications that could not be indexed by a traditional search engine [4]. While the Dark Web may host legitimate content and services, it is also notorious for illicit activities, including the sale of drugs, weapons, stolen data, and hacking tools [13]. It has also become a hub for various illegal marketplaces, forums, and communities, making it a subject of concern for law enforcement agencies worldwide.

The dark side of the Metaverse could therefore be defined as the Darkverse, and involve illicit activities taking place within a virtual space, mirroring those already occurring on the Dark Web and physically in reality. Challenges of the Darkverse could similarly include user vulnerability, privacy, identity theft, invasive advertising, misinformation and propaganda, phishing, fraud, illicit activities, abuse, pornography, social exclusion, sexual harassment and other unintended consequences [7].

Potential **research challenges** can thus be summarized as follows: **RC1**: What similar and new forms of illicit activities can be enabled within such virtual and embodied spaces? **RC2**: How are such illicit and criminal activities detected and identified within the Darkverse? **RC3**: How are such illicit activities mitigated and reduced in such an environment?

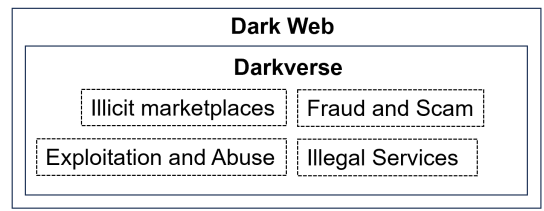

Figure 1: Criminal activities in Darkverse.

The subsequent sections discuss potential criminal activities enabled by the Darkverse (Section 2), and the similarities between the Darkverse and Dark Web are compared (Section 3). The challenges of identifying criminal activities in virtual environment are discussed (Section 4) based on the existing issues in Dark Web. Finally, potential directions and future work are explored.

## 2 CRIMINAL ACTIVITIES IN DARKVERSE

In this section, we introduce some common criminal activities on the Dark Web and discuss how they could be enabled in the Darkverse:

**Illicit marketplaces**: Dark Web marketplaces serve as platforms for buying and selling illicit drugs, firearms, explosives, and other weapons [12]. These marketplaces often operate similarly to e-commerce platforms, with sellers offering various drugs for sale and buyers making purchases using cryptocurrencies. While regulations and restrictions on weapons vary by jurisdiction, the anonymity of the Dark Web makes it easier for individuals to circumvent these laws and acquire weapons illegally.

The Darkverse may emerge as a new platform for criminals to trade drugs and weapons. Criminals could create virtual versions of these items to attract potential buyers and provide proof of possession for the items.

**Fraud and Scams**: Dark Web forums and marketplaces are used for buying and selling stolen personal information, such as credit card numbers, bank account details, and Social Security numbers [18].

In addition to trading personal information, criminals may also exploit the Darkverse for phishing and scams by using the acquired personal information [5].

**Exploitation and Abuse**: The Dark Web is known for hosting websites and forums dedicated to the distribution of child pornography and the facilitation of child sexual exploitation. Law enforcement agencies around the world actively monitor these sites and work to identify and prosecute individuals involved in these heinous crimes [9].

**Illegal Services**: Beyond goods, the Dark Web offers illegal services such as hacking services, hitman-for-hire services, and various forms of fraud assistance. It is also a marketplace for buying and selling hacking tools, exploit kits, and malware.

These tools can be used to compromise computer systems, steal data, and launch cyber attacks against individuals, businesses, and government agencies.

It is likely that the criminal activities mentioned above will occur in Darkverse. Therefore, Darkverse will become one of the most serious security concerns that law enforcement may need to investigate.

## 3 SIMILARITIES BETWEEN DARK WEB AND DARKVERSE

In this section, the technologies between the Dark Web and the Darkverse and compared.

(1) Search capability: It is difficult to find the generated content inside the Darkverse application as the existing search engines has not developed any crawling technologies to index the content inside Darkverse.
(2) Anonymous characteristics: Users can generate their own characters in Darkverse which may not be able for people to identify the real person. Similar to famous virtual YouTuber nowadays. Criminal may use technology like DeekFake to make even harder to identify the real person.
(3) In depth understanding in technology: Normal user might not be able to discover and access Dark Web if they do not have the knowledge. Darkverse may increase the challenge even further as it requires a XR headset to connect with. The user may need to modify the setting of the headset to browse Darkverse content.

## 4 CHALLENGES OF IDENTIFYING CRIMINAL ACTIVITIES IN DARKVERSE

In this section, we discuss the possible challenges to the identification and tracing of illicit activity in the Darkverse [20].

(1) No recording and logging in the application. As there are huge amount of data as to be processed in the Darkverse application.
(2) No personally identifiable information is sometimes captured. Criminals can use the same avatar and voice to communicate with each other.
(3) Flash mob activities. Unlike the usual way that criminals discuss in the Dark Web forums, they can just pop-up in any area in the Darkverse and communicate. It is unlikely to find traces of their activities afterwards.

## 5 CURRENT AND FUTURE WORK

The authors are currently working together with the local government and local law enforcement agencies to identify potential new risks and harms associated with the Metaverse, exploring technology platforms for the mitigation of these risks and also further understanding negative user and social behaviors within the virtual environment.

The team is proposing to create a testbed based on OpenXR [10], which allows stakeholders and policymakers to flexibly and effortlessly monitor user activities and evaluate various trust frameworks, verification methods and policies within the virtual environment. Examples of which are methods for user identification, privacy-preserving frameworks for capture and retention of data, and aggregated data monitoring methods.

## 6 CONCLUSION

The Darkverse is introduced, mirroring the Dark Web, namely application platforms facilitating illicit activities occurring within the virtual space of the Meteaverse. The possible criminal activities that could occur within the Darkverse are enumerated and compared, together with challenges for identification of such crimes. Future directions of user evaluation, development of a testbed and creation of privacy preserving frameworks are then discussed. It is hoped that new strategies and mitigation methods could be developed to reduce the harms in the future Metaverse for users.